\documentclass{iucrjournals}

\usepackage{multirow}
\usepackage{booktabs}
\usepackage{amsmath}

\title{Application of X-ray Absorption Spectroscopy with a laboratory X-ray powder diffractometer}

\author{Milen Gateshki\IUCrCemaillink{milen.gateshki@malvernpanalytical.com}\IUCrOrcidlink{0000-0003-3251-6266}}%
\author{Charalampos Zarkadas\IUCrEmaillink{charalampos.zarkadas@malvernpanalytical.com}\IUCrOrcidlink{0009-0001-6690-0533}}%
\author{Detlef Beckers\IUCrEmaillink{detlef.beckers@malvernpanalytical.com}\IUCrOrcidlink{0009-0005-6774-9408}}%

\affil{Malvern Panalytical B. V., Lelyweg 1, 7602 EA, Almelo, The Netherlands}

\begin{document} 
\maketitle 

\begin{synopsis}
A method for high-quality X-ray absorption spectroscopy (XAS) in transmission mode is proposed that can be easily applied on a laboratory X-ray powder diffractometer. Such application may contribute to the wider use of this powerful technique in materials research laboratories.
\end{synopsis}

\begin{abstract}
A novel method and experimental configuration are proposed that allow the collection of high-quality X-ray absorption spectroscopy (XAS) data in transmission mode on a standard laboratory diffractometer. This configuration makes use of components that have been developed for application in modern laboratory diffractometers, such as X-ray tubes with enhanced stability, accurate goniometers and position-sensitive photon-counting detectors with improved energy resolution. Such approach may extend the application of XAS in materials research, particularly for users of diffraction methods. 
\end{abstract}

\keywords{Laboratory XAS; XANES; EXAFS; X-ray diffraction; In operando studies }

\section{Introduction}

XAS is a powerful method for analyzing the structure and properties of new materials. It investigates the behavior of the absorption coefficient around the absorption edge of a specific element. The measured absorption spectrum is usually divided into two ranges: XANES (X-ray absorption near edge structure), which is up to 50–100 eV above the absorption edge of the investigated atomic type and provides information about the oxidation state, electronic structure and local environment of the atom, and EXAFS (Extended X-ray absorption fine structure) that extends from 100 to 500-2000 eV above the edge and probes the short-range ordering of the atoms, i.e. interatomic distances and coordination numbers. XAS provides complementary information to X-ray diffraction (XRD) and X-ray fluorescence (XRF) and can be used for relatively complex materials where other techniques are not applicable, e.g. liquids and amorphous substances. Compared to the Pair Distribution Function method (PDF), which is also applied to study short-range ordering in materials (including amorphous) and gives averaged information for all atomic species, XAS is element specific and can be used to distinguish between the environments of otherwise similar atoms, e.g. two types of metal atoms in an alloy or a solid solution.  

During the last decade, the interest in laboratory-based XAS methods has continually grown both in the academic and in the industrial sectors \cite{Zimmermann2020, Cutsail2022}. Nowadays, several companies, such as easyXAFS, Sigray, HP Scpectroscopy, LynXes, RapidXAFS and Specreation offer a range of dedicated XAS instruments. 
Two types of experimental set-ups are most commonly found in XAS instruments: the Rowland circle geometry and the von Hámos geometry. Both focus the X-ray beam from the source to the position of the detector. The Rowland circle geometry uses a cylindrically or spherically bent crystal analyzer \cite{Mottram2020, Lutz2020, Jahrman2019a, Holden2017, Seidler2014, Seidler2016, Honkanen2019, Honkanen2014}, while the von Hámos geometry employs a crystal that is cylindrically bent in transverse direction \cite{Hamos1932, Hamos1933, Schlesiger2020, Nemeth2016, Szlachetko2012, Zeeshan2019, Alonso-Mori2012, Ismail2021}. Detailed discussions of these geometries and their corresponding advantages in XAS experiments can be found in the referenced publications. Focusing geometries can significantly increase the intensity of the measured signal, however they require instruments with increased mechanical complexity. Moreover, they are not compatible with the architecture of laboratory X-ray powder diffractometers, which are based on highly accurate goniometers with the X-ray source mounted on one arm of the goniometer and the X-ray detector mounted on the second arm.

\section{XAS measurement configuration} \label{sec:section2}

\begin{figure}[ht] %
\begin{center}
\includegraphics[width=0.5\textwidth]{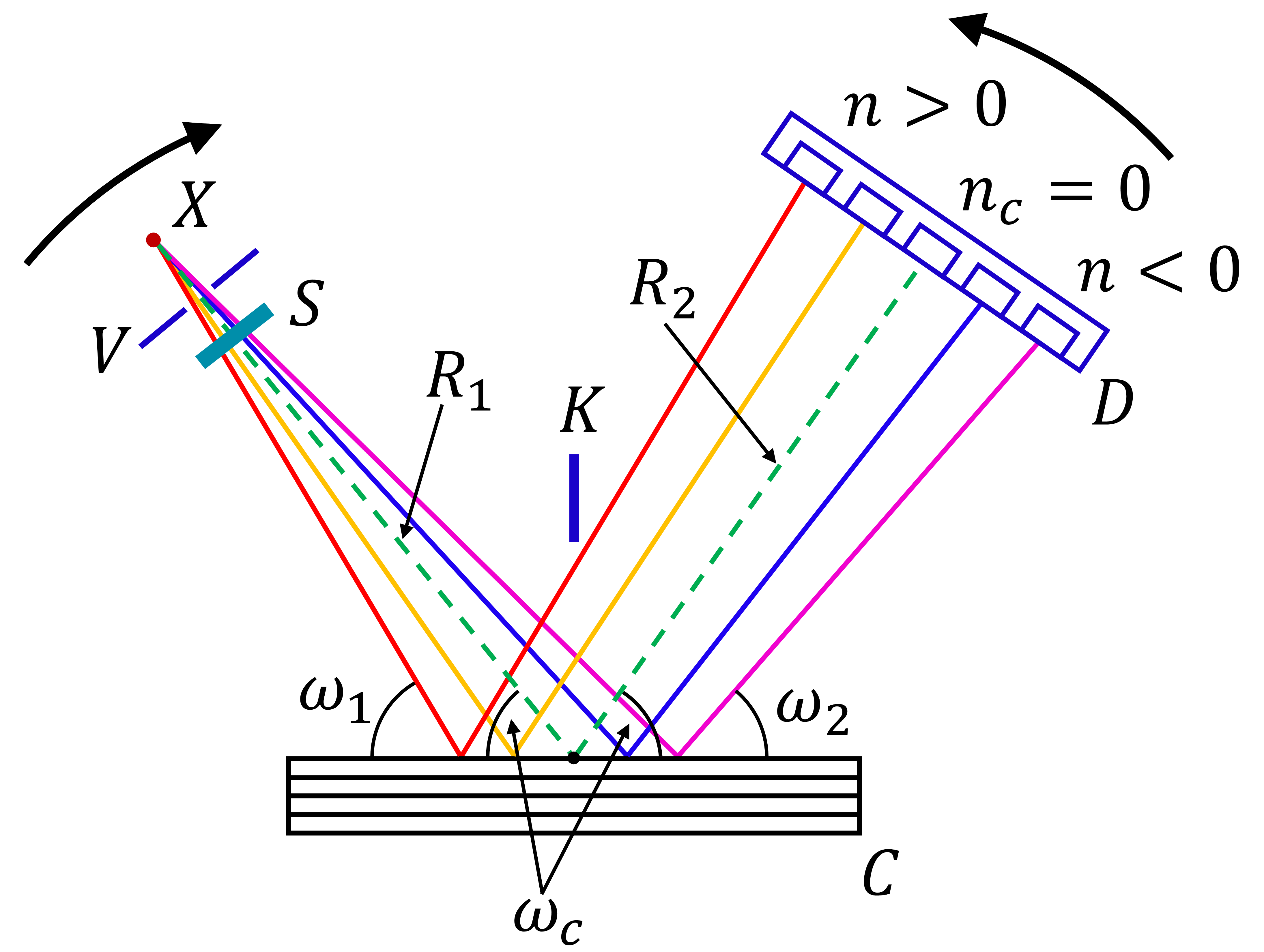}
\end{center}
\caption{A schematic diagram of the experimental set-up. Polychromatic X-ray radiation is emitted by the X-ray source $X$, travels through specimen $S$ and is diffracted by the flat crystal monochromator $C$. After diffraction, the X-ray photons are directed to the position sensitive detector $D$. Depending on the angle of incidence, photons with different energies will be diffracted at different points on the crystal surface. This is represented by lines of different colors. $R_1$ and $R_2$ are the distances between the source and the rotation axis (center) of the goniometer and between the goniometer center and the middle channel of the X-ray detector. The top surface of the crystal is aligned to be at the center of the goniometer. $V$ is a divergence slit, and $K$ is a beam shield.}
\label{fig:figure1}
\end{figure}

In this work, we propose a configuration for XAS that uses a flat analyzer crystal and can be easily applied on a standard laboratory X-ray powder diffractometer. A schematic representation of this set-up is shown in Fig. \ref{fig:figure1}. The figure gives a side view of the configuration, which uses an XRD tube with a long fine focus (LFF) and an one-dimensional (strip) XRD detector with high energy resolution, e.g. better than 500 eV. The energy resolution of the detector helps to reduce background effects, such as higher harmonics, characteristic lines from the X-ray tube, fluorescence, etc.  The divergence slit is used to adjust the size of the incident beam and several antiscatter devices can be introduced to reduce scattered radiation from optical components or from the air. The axial divergence of the incident and diffracted beams can be controlled with Soller slits, which are also included in the basic configuration of laboratory powder diffractometers.  The sample holder is attached to the goniometer arm and it moves together with the X-ray tube. In some cases it may be preferable to attach the sample holder to the second arm and to move it together with the detector. The same geometry can be applied also with a microfocus tube and a two-dimensional (area) detector. In this case the use of Soller slits is not required. 
Since powder diffraction measurements are most often performed in air environment, the XAS experiments presented in the following sections were also conducted in air, although a configuration that uses an evacuated or helium-filled beam path can be also envisioned. 

Fig. \ref{fig:figure1} depicts a wavelength-dispersive configuration, in which photons originating from the focus of the X-ray tube $X$ travel through sample $S$ and impinge on the surface of crystal $C$ at different angles. Following Bragg’s law of diffraction, only X-ray photons with specific energies will be diffracted by the crystal towards the X-ray detector $D$. Since the incident photons shown in Fig. \ref{fig:figure1} form different angles with the atomic planes of the crystal, each one of the diffracted beams (shown as lines of different colors in Fig. \ref{fig:figure1}) will have a different wavelength. Therefore, the one-dimensional detector collects simultaneously a range of different energies. This can significantly increase the speed of the measurement compared to a set-up that uses a single-element (point) detector. Considering the exact geometry of the experimental set-up, it is possible to correlate the position (strip number), at which the signal is received, with the energy of the diffracted beam. In this way, a complete energy spectrum of the X-rays transmitted through the sample can be obtained and the absorption at each energy can be determined. 

\begin{figure}[ht!] %
\begin{center}
\includegraphics[width=0.5\textwidth]{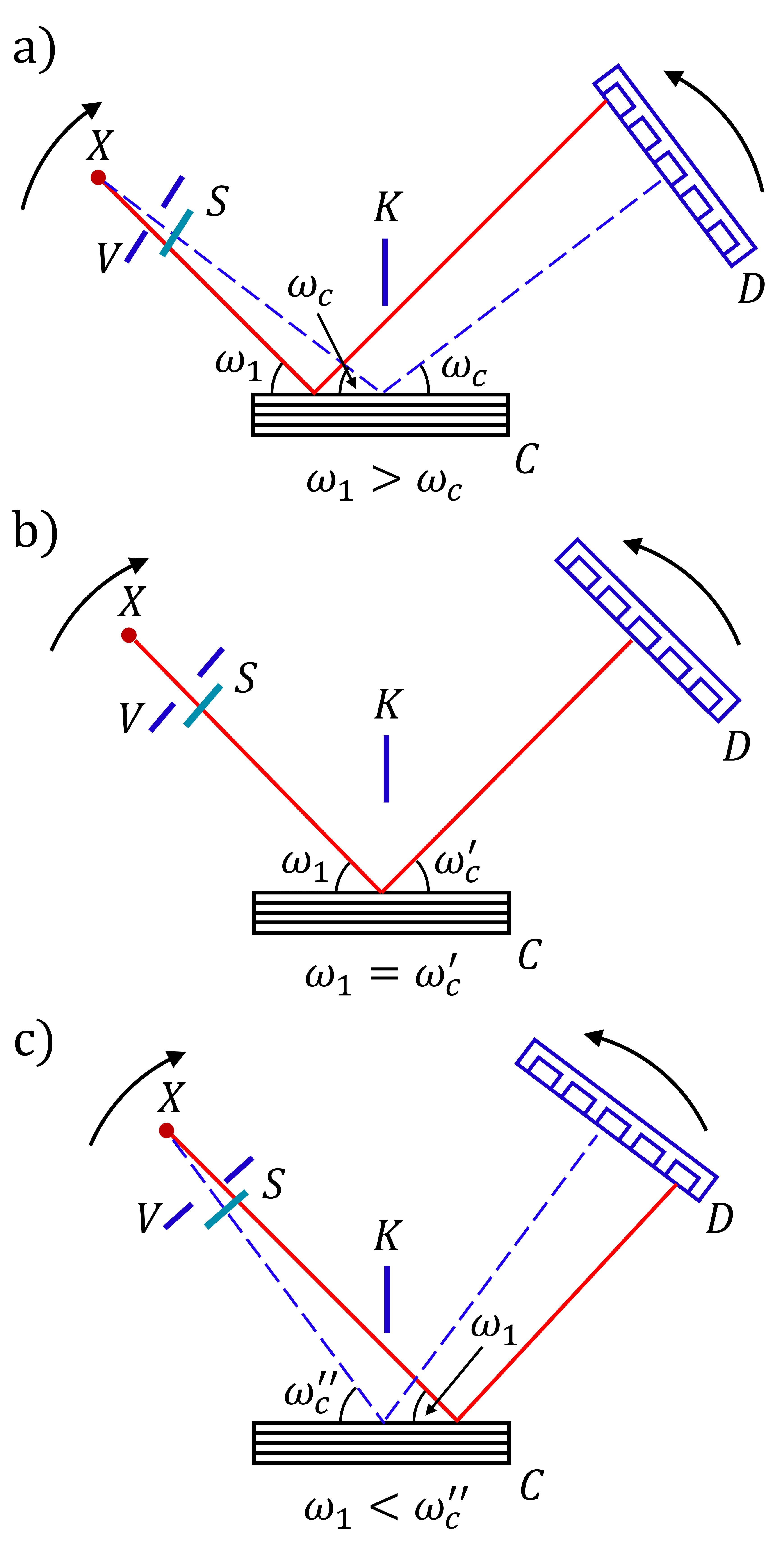}
\end{center}
\caption{Schematic representation of XAS measurements with the configuration shown in Fig. \ref{fig:figure1} and using a continuous scanning movement of the X-ray source and detector relative to the crystal. (a) Lower angle of incidence of the central part of the X-ray beam, (b) intermediate and (c) higher angle of incidence. The notation for the different elements is the same as in Fig. \ref{fig:figure1}. } 
\label{fig:figure2}
\end{figure}

The range of energies that can be collected simultaneously depends on the size and position of the detector. If the covered angular range is narrower than the intended one (either XANES or EXAFS), it can be extended by operating the detector in scanning mode with the goniometer arm moving continuously during the measurement from low to high diffraction angles. Note that maintaining the diffraction condition requires that the X-ray source moves with the same speed as the detector. This measurement mode (known as gonio mode or symmetric scanning) is a standard feature of X-ray diffractometers and does not require special modifications of the hardware or software. In addition to the extended range, the scanning mode has additional advantages for XAS measurements compared to the static mode. It enables collecting data with a step size that is smaller than the angular size of one detector strip (subsampling), which provides better discretization of the continuous XAS signal and allows the observation of narrow spectral features. A second advantage is that during the scanning movement, X-ray beams with the same photon energy will travel through different parts of the sample, thus effectively averaging the signal. In this sense, the resulting measurements will be less sensitive to small variations of the sample’s composition or thickness. This effect is illustrated in Fig. \ref{fig:figure2}. At lower positions of the X-ray source and detector, an X-ray beam passes through the lower part of the sample (Fig. \ref{fig:figure2}(a)) and is detected at the upper end of the detector. As the source and detector move higher (Fig. \ref{fig:figure2}(b)), a beam with the same photon energy (same angle of incidence $\omega_1$ on the crystal) will pass through the central part of the sample and will be detected in the center of the detector. Finally, at even higher positions of the source and detector (Fig. \ref{fig:figure2}(c)), photons with once again the same energy will travel through the top part of the sample and will be detected at the lower edge of the detector. 
From Fig. \ref{fig:figure2} it can be also observed that for each position of the X-ray source, the photons with the specified energy will be diffracted by a different point on the crystal. As a result of this, local variations in the quality of the crystal surface will be also averaged during the scanning measurement. 

The data collection algorithm in the detector combines all intensity that is received for a given angle of incidence and produces a single data point that corresponds to this angle of incidence (and hence photon energy). Considering the configuration shown in Fig. \ref{fig:figure1}, the angle of incidence for a specific photon direction can be calculated from the number of the detector channel in which the signal is received by means of equation (\ref{eq:eq1}). For derivation, see Fig. S1. $R_1$ is the distance between the X-ray source and the center of the goniometer, $R_2$ is the distance between the goniometer center and the detector, n is the number of the detector strip counted from the middle channel of the detector, p is the width of the strip and $\omega_c$ is the angle between the central part of the incident beam (indicated with a dashed line in the figure) and the diffracting planes of the crystal monochromator.
 
 \begin{equation}
		 \omega = \omega_c + \tan^{-1} \left[\frac{np}{R_1+R_2}\right] \approx \omega_c + \frac{np}{R_1+R_2}
 \label{eq:eq1}
 \end{equation} 

\begin{figure}[ht] %
\begin{center}
\includegraphics[width=0.5\textwidth]{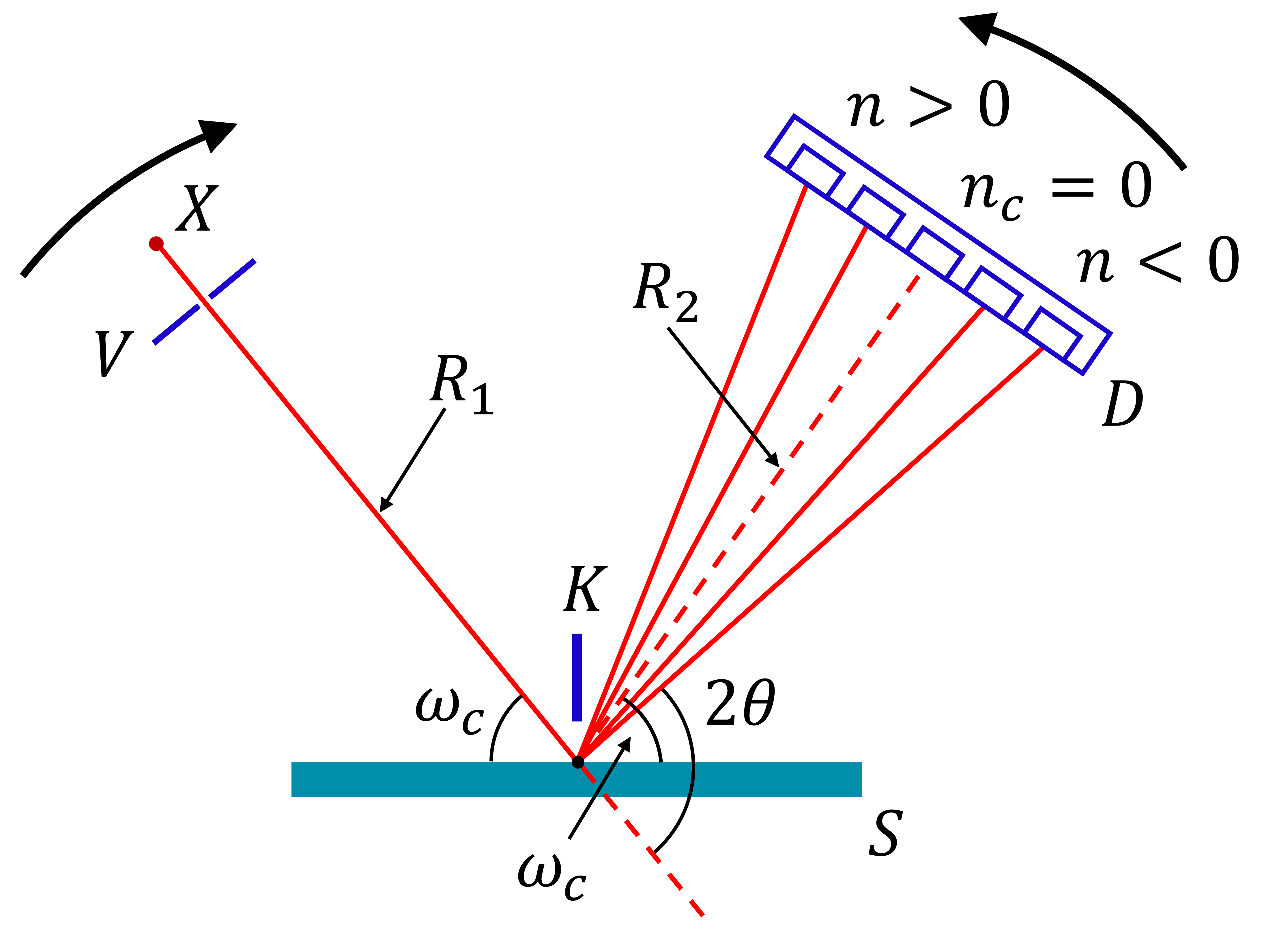}
\end{center}
\caption{Schematic representation of X-ray powder diffraction measurements with a narrow incident beam, a one-dimensional position-sensitive detector and a symmetric (gonio) scan. The notation for the different elements is the same as in Fig. \ref{fig:figure1}.}
\label{fig:figure3}
\end{figure}

The measurement method described by Fig. \ref{fig:figure1} and equation (\ref{eq:eq1}) is different from the one that is used in powder diffraction measurements. A diagram of a diffraction measurement with monochromatic radiation, a symmetric (gonio) scan and a narrow incident beam is shown in Fig. \ref{fig:figure3}. In this method, the algorithm for processing data from the detector is set up for measuring the intensity at each diffraction angle 2$\theta$. If it is assumed that $\omega$ = 2$\theta$/2, equation (\ref{eq:eq2}) can be derived. From this result it can be observed that for $R_1 = R_2$ the approximated expressions for $\omega$ are the same in equations (\ref{eq:eq1}) and (\ref{eq:eq2}). Therefore, the data collection algorithm for powder diffraction measurements with symmetric scans, which implements equation (\ref{eq:eq2}), can be applied also for XAS without modifications. If $R_1 \neq R_2$, then the distance $R_2$ must be replaced in (\ref{eq:eq2}) by ($R_1$ + $R_2$)/2. Equation \ref{eq:eq2} is also applied for XRD measurements with the Bragg-Brentano parafocusing geometry, which uses a divergent monochromatic incident beam and is the most commonly used type of measurement in laboratory powder diffractometers.         

 \begin{equation}
		 \omega = \frac{2\theta}{2} = \omega_c + \frac{1}{2}\tan^{-1} \left[\frac{np}{R_2}\right] \approx \omega_c + \frac{np}{2R_2}
 \label{eq:eq2}
 \end{equation} 

A configuration somewhat similar to the one shown in Fig. \ref{fig:figure1} is reported in \cite{Nemeth2016}, where flat sections of Si crystal are arranged around a cylindrical surface to approximate the von Hámos geometry. In this set-up, both X-ray source and detector are located on the axis of the cylinder. The detector is mounted on a translation stage, instead of a goniometer arm, and the X-ray source is in the plane of the detector, which in turn is not perpendicular to the direction of the X-ray beams diffracted by the crystal analyzer. The measurements are performed in static mode, which does not make use of the advantages associated with scanning measurements that were described in the previous paragraph. On the other hand, the set-up in \cite{Nemeth2016} uses distances between the main components that are similar to the ones reported in this work, and experiments are performed in air atmosphere. This indicates that a compact configuration operated in air is a feasible solution for XAS experiments. 

\section{Experimental details}

The experimental configuration described in Section \ref{sec:section2} was realized on an Empyrean multipurpose diffractometer (Malvern Panalytical B.V., Almelo, The Netherlands) with the following components: 
\begin{itemize}
\item	Empyrean long fine focus X-ray tube with enhanced stability of the focal spot (LFF HR) and either Cu or Ag anode. The take-off angle is 6 deg and the apparent size of the focus is 0.04 by 12 mm.
\item	Fixed divergence slit with opening either 1 deg or $1/4$ deg, two 0.04 rad Soller slits for controlling the axial divergence of the incident and diffracted beams, and a 11.6 mm mask for limiting the beam size in axial direction. With these settings the irradiated area on the specimen is about 12 mm by 1 mm (1 deg slit) or 12 mm by 0.25 mm ($1/4$ deg slit). If it is required to reduce the lateral size of the irradiated area, this can be done by using a smaller mask.
\item	Flat Si (011) crystal with dimensions 40 mm (length) by 20 mm (width). The top surface of the crystal is aligned to be at the center of the goniometer.
\item	1Der (2\textsuperscript{nd} generation) one-dimensional solid-state X-ray detector that consists of 127 strips with dimensions 0.07 mm wide and 15 mm long. The energy resolution of the detector is 320 eV, which is considered sufficient for XAS experiments \cite{Welter2009}.
\item	For the XAS experiments reported in this work, the distances $R_1$ and $R_2$ were both equal to 240 mm.
\end{itemize}

\subsection{Choice of anode material}

Modern diffractometers offer the option for easy exchange of X-ray tubes and this opens the possibility for selecting a tube with an anode material and power settings that are best suited for a specific material. When selecting an X-ray tube for XAS experiments, it is important to consider the intensity of the $bremsstrahlung$ radiation emitted from the tube in an energy range close to the absorption edge of the element of interest. To evaluate this intensity for X-ray diffraction LFF tubes, the continuum radiation of various commonly used anode materials was calculated at different voltage settings by means of a tube spectrum model widely applied in X-ray spectrometry \cite{Ebel1999, Ebel2005}. The results of the theoretical calculations are summarized in Table 1. From the values presented there it becomes clear that the performance of a tube drops when the absorption edge of the investigated element is at a higher energy compared to the absorption edge of the anode element. This is due to the effect of anode self-absorption that reduces the absolute number of continuum photons available for XAS measurements. One remedy for this is to use a different tube, with an anode material which is more favorable. In cases where materials with the same absorption edge are analyzed repeatedly, the optimal anode material can be selected from Table \ref{tab:table1}. Conversely, in a situation where different absorption edges are probed, many materials of interest can be analyzed using either Cu or Ag tubes with only a small reduction in performance. Using X-ray tubes with different anode materials may also help to avoid characteristic lines that appear in the measurement range in specific cases. Such characteristic lines with high intensities and narrow peak widths can disturb the normalization procedure and produce artifacts in the data.


\begin{table}[ht]
\caption{Optimal choices of anode materials and tube voltage settings for XANES measurements around different absorption edges. The corresponding tube current (not shown in the table) corresponds to isowatt settings at all tabulated voltages. The maximum power used for the calculations was 1.8 kW for Co, Cu and Ag tubes and 2.5 kW for Mo tubes.} 
\label{tab:table1}
\smallskip
\begin{center}

\begin{tabular}{p{2cm} p{1.5cm} |p{1.5cm} p{1.5cm}| p{1.5cm} p{2cm}| p{1.5cm} p{2cm} }
& & \multicolumn{2}{|c}{Optimal selection} & \multicolumn{2}{|c}{Cu tube} & \multicolumn{2}{|c}{Ag tube}  \\
\midrule
 Absorption edge    & Energy, keV        &  Anode material   &    Tube voltage, kV  &    Tube voltage, kV &    Intensity compared to optimal selection, \% &    Tube voltage, kV & Intensity compared to optimal selection, \%   \\
\midrule
 Ti      & 4.966      & Co      & 30   & 30      & 93.6      & 30   & 38.7 \\
 Cr      & 5.989      & Co      & 35   & 30      & 97.8      & 30   & 50.1 \\
 Fe      & 7.112      & Co      & 35   & 35      & 99.4      & 30   & 63.5 \\
 Ni      & 8.333      & Cu      & 35   & 35      & 100.0     & 30   & 77.4 \\
 Cu      & 8.979      & Mo      & 45   & 30      & 34.9      & 30   & 85.0 \\
 Pt L$_3$ & 11.564    & Mo      & 45   & 30      & 34.8      & 35   & 76.8 \\
 Se      & 12.658     & Mo      & 45   & 35      & 35.8      & 35   & 75.1 \\
 Sr      & 16.105     & Mo      & 55   & 40      & 37.2      & 45   & 72.5 \\
 Zr      & 17.998     & Mo      & 60   & 45      & 37.6      & 55   & 72.4 \\
 Mo      & 20.000     & Ag      & 60   & 50      & 52.1      & 60   & 100.0\\
 Pd      & 24.350     & Ag      & 60   & 60      & 55.5      & 60   & 100.0\\
 Ag      & 25.514     & Mo      & 55   & 60      & 68.4      & 50   & 71.9 \\
\end{tabular}

\end{center}
\end{table}

\subsection{Evaluation of the energy resolution}

\begin{figure}[ht] %
\begin{center}
\includegraphics[width=0.5\textwidth]{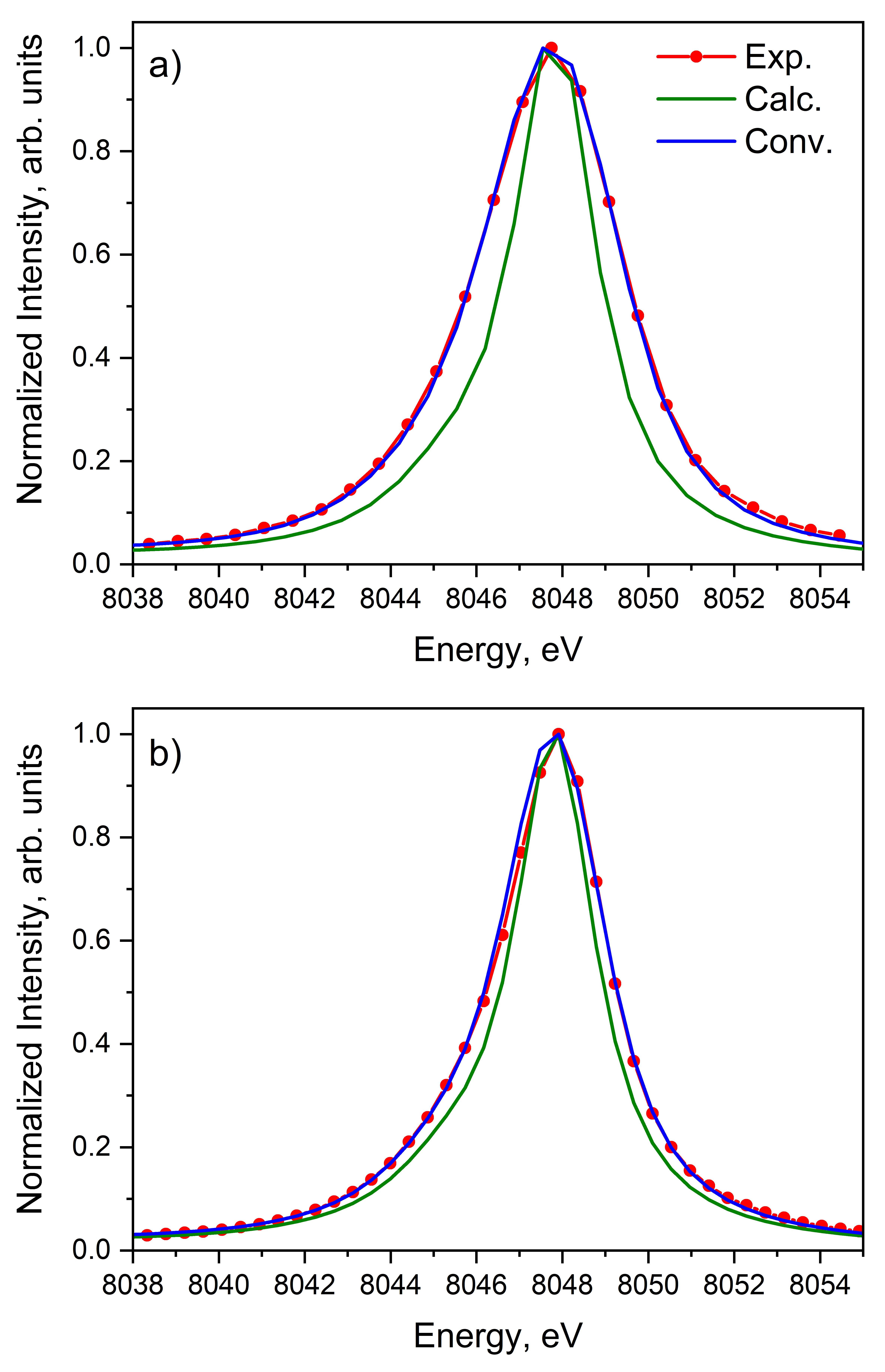}
\end{center}
\caption{Experimental profile of the Cu$K\alpha_1$ characteristic line (red dots), compared with the expected theoretical profile (green line) and the theoretical profile convoluted with Gaussian instrumental broadening (blue line). The estimated instrumental broadening is equal to 2.2 eV for the Si (022) reflection (a) and 1.3 eV for Si (044) (b).}
\label{fig:figure4}
\end{figure}

One of the most important criteria that determines the quality of XAS experiments is the achievable energy resolution, particularly in the XANES region. The resolution is determined by multiple factors, such as geometrical parameters (distances from the crystal to the source and the detector, widths of the X-ray source and detector strips) as well as factors related to the crystal, such as flatness, quality of the surface and the Darwin width of the selected reflection. To evaluate the combined energy resolution resulting from all experimental factors, the profile of the Cu$K\alpha_1$ emission line from the X-ray source was measured using the Si (022) reflection of the crystal and compared (Fig. \ref{fig:figure4}(a)) with a calculated profile based on tabulated values for the widths of the different spectral components \cite{Deutsch2004}. In this way, it was determined that the energy resolution of the experimental setup described above was $\sim$2.2 eV at the energy of 8 keV, resulting in resolving power of $E/{\delta}E = 3600$. This resolution is adequate for many XANES experiments \cite{Nemeth2016, Schlesiger2020}. If higher resolution is required, then a different reflection or a different crystal could be employed. For example, the Si (044) reflection of the same crystal can be used (Fig. \ref{fig:figure4}(b)), which gives resolution of $\sim$1.3 eV at 8 keV ($E/{\delta}E = 6150$). This improved resolution comes at the cost of $\sim$10 times reduction in intensity and correspondingly longer measurement times.
Due to the specifics of the method, it is expected that the absolute energy resolution will gradually improve at energies lower than 8 keV and become worse at higher energies.

\subsection{Effect of air absorption }

\begin{figure}[ht] %
\begin{center}
\includegraphics[width=0.5\textwidth]{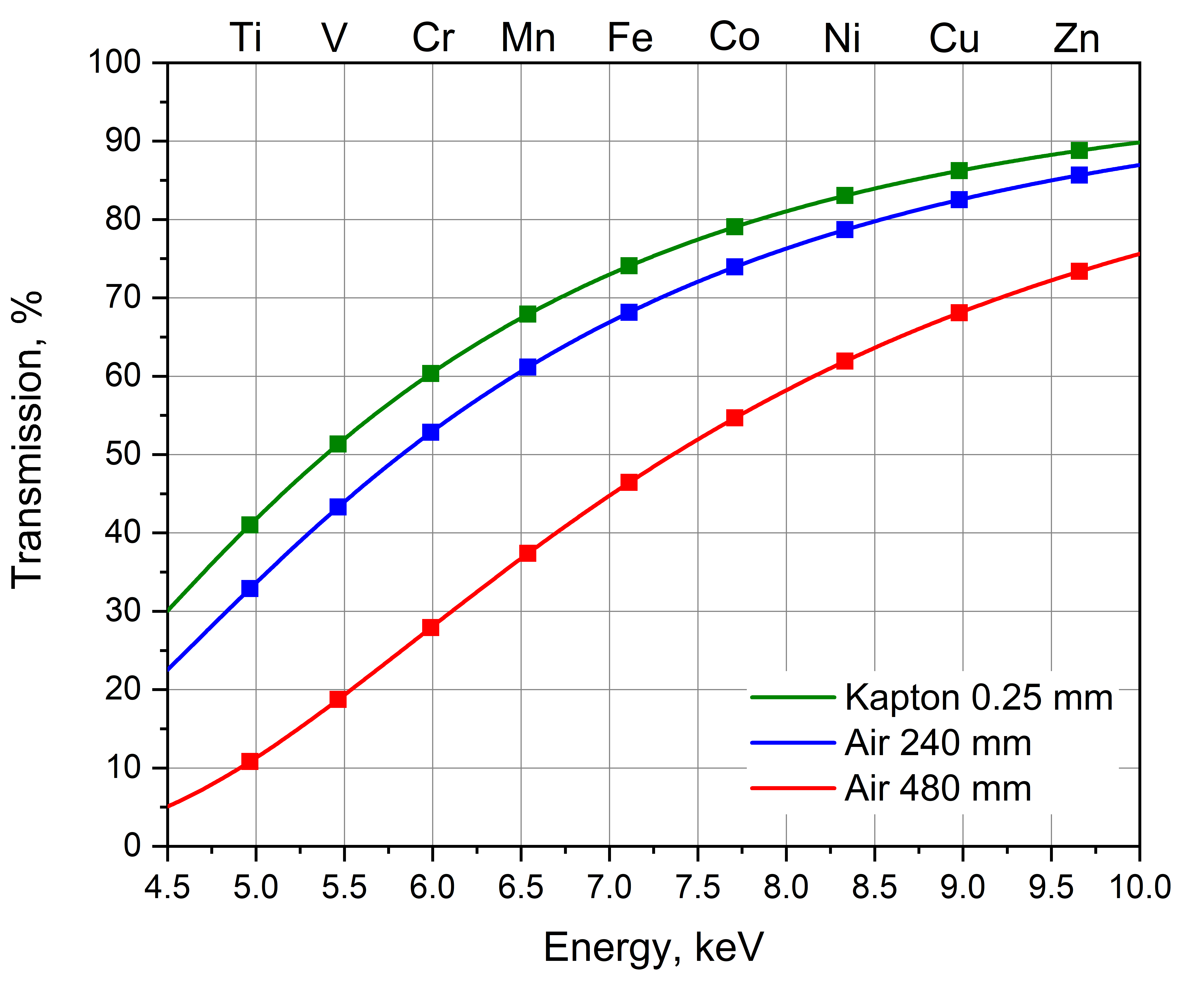}
\end{center}
\caption{Transmission coefficient for X-rays corresponding to the absorption edges of different materials calculated for beam paths of 240 mm dry air (blue), 480 mm dry air (red) and Kapton (polyimide) windows with 0.25 mm total thickness (green).} 
\label{fig:figure5}
\end{figure}

Absorption of the transmitted X-ray photons by air can lead to a significant reduction of measured intensities, particularly in the low-energy range. Because of this, XAS instruments often use evacuated or He-filled chambers. Such implementation, however, limits the flexibility of the set-up and is difficult to integrate with the diffractometer framework. The effect of air absorption for the configuration discussed in this work was evaluated and the results are shown in Fig. \ref{fig:figure5}. For the energies above 8 keV, the transmission through 480 mm of air is about 80\% and the effect of air absorption on the experimental data will be barely noticeable. On the other hand, in the range around 5 keV, the transmission is only about 10\%. Such intensity reduction will require longer measurement times. One possibility to improve this is to select an X-ray tube and tube power settings that optimize the emission of X-rays with such energies (see Table \ref{tab:table1}). If the use of an evacuated chamber is desired for reducing air absorption, it should be considered that for the proposed implementation, such chamber can cover only a part of the beam path, as the X-ray tube, sample, and detector will remain outside of the chamber. In addition, the X-ray beam will have to traverse either two or four chamber windows, depending on whether the crystal analyzer is inside or outside of the chamber. 

\section{Results and discussion}

The performance of the set-up described in Section \ref{sec:section2} is demonstrated by comparing measured XAS spectra of reference metal foils (supplied by EXAFS materials, Danville, CA, USA) with high-quality synchrotron data obtained from the Materials Data Repository (https://mdr.nims.go.jp/). Measurements of metal oxides were also conducted with samples prepared by mixing the corresponding powder materials with isopropyl alcohol and depositing them on a Kapton (polyimide) foil. The examples discussed further in this section were selected to emphasize different aspects of the XAS data collection, such as measurement time, resolution, energy range and fidelity of the measured data and also to illustrate the application of the set-up to practical cases. 
The collected XAS spectra were processed using the program $Athena$ \cite{Ravel2005}.

\subsection{Fe foil, $\alpha$-Fe$_2$O$_3$ and Fe$_3$O$_4$}

\begin{figure}[ht] %
\begin{center}
\includegraphics[width=0.5\textwidth]{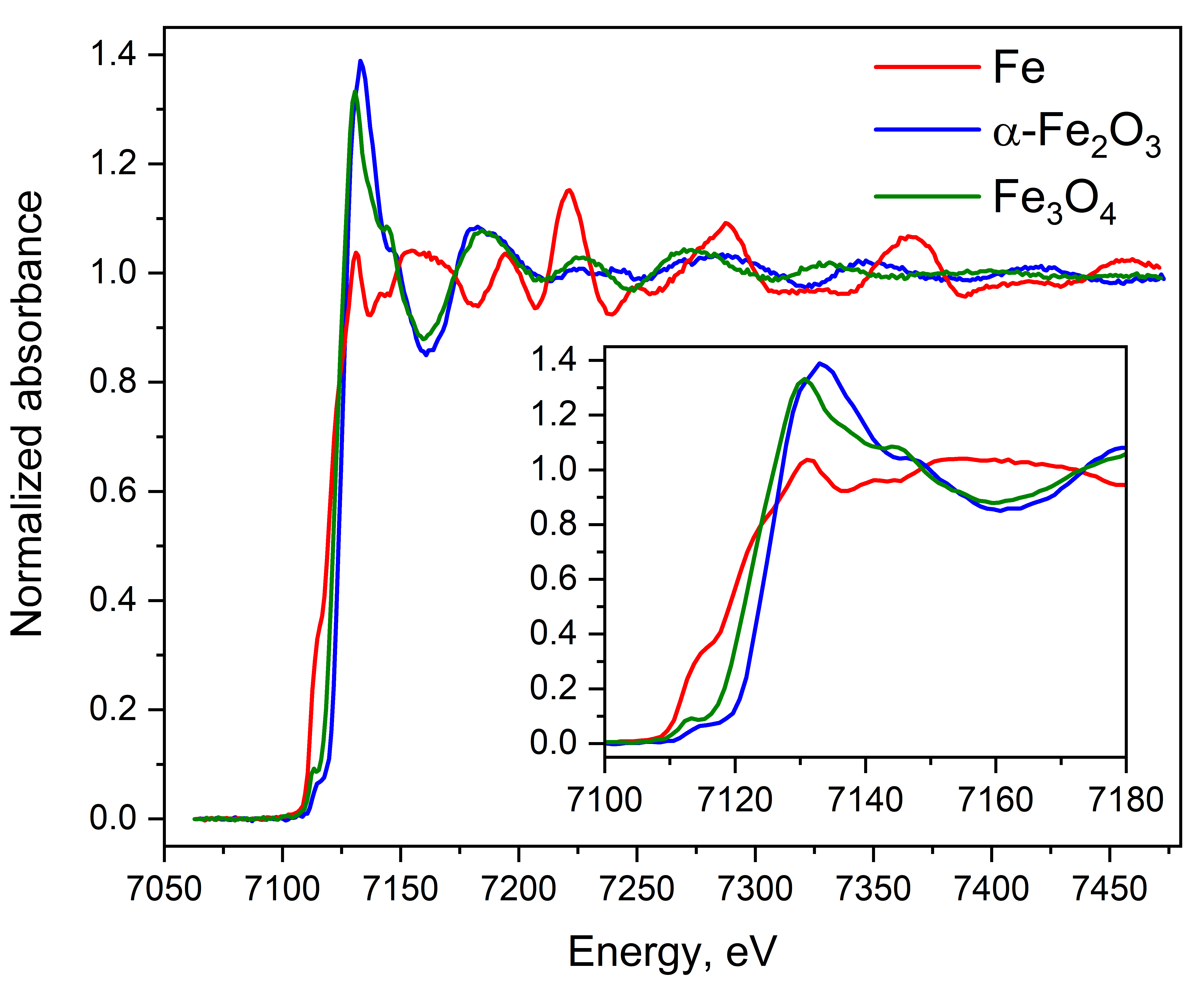}
\end{center}
\caption{XAS spectra of $\alpha$-Fe$_2$O$_3$ (blue) and Fe$_3$O$_4$ (green) powder samples, compared with the corresponding spectrum of a 7.5 $\mu$m Fe reference foil (red). The inset shows a magnified view of the XANES range.} 
\label{fig:figure6}
\end{figure}

The XAS spectra of $\alpha$-Fe$_2$O$_3$ (hematite) and Fe$_3$O$_4$ (magnetite) samples prepared from powder are shown in Fig. \ref{fig:figure6} and are compared with the corresponding spectrum of a Fe foil with a thickness of 7.5 $\mu$m. Following Table \ref{tab:table1}, a Cu tube with power settings 35 kV and 50 mA was applied. A divergence slit of $1/4$ deg was used and the measurement time for each of the three scans was one hour. A scan of equal duration and with the sample removed from the beam path was also collected for normalization of the data (see Fig. S2). The differences in the positions and shapes of the absorption edge of Fe are clearly observed, indicating the different oxidation states and coordinations of the Fe atoms in the three materials. The pre-edge effects in the oxides are also visible. The results are consistent with previous reports based on synchrotron studies, e.g. \cite{Xiao2022}. 

\subsection{Ni foil and NiO}

\begin{figure}[ht] %
\begin{center}
\includegraphics[width=0.5\textwidth]{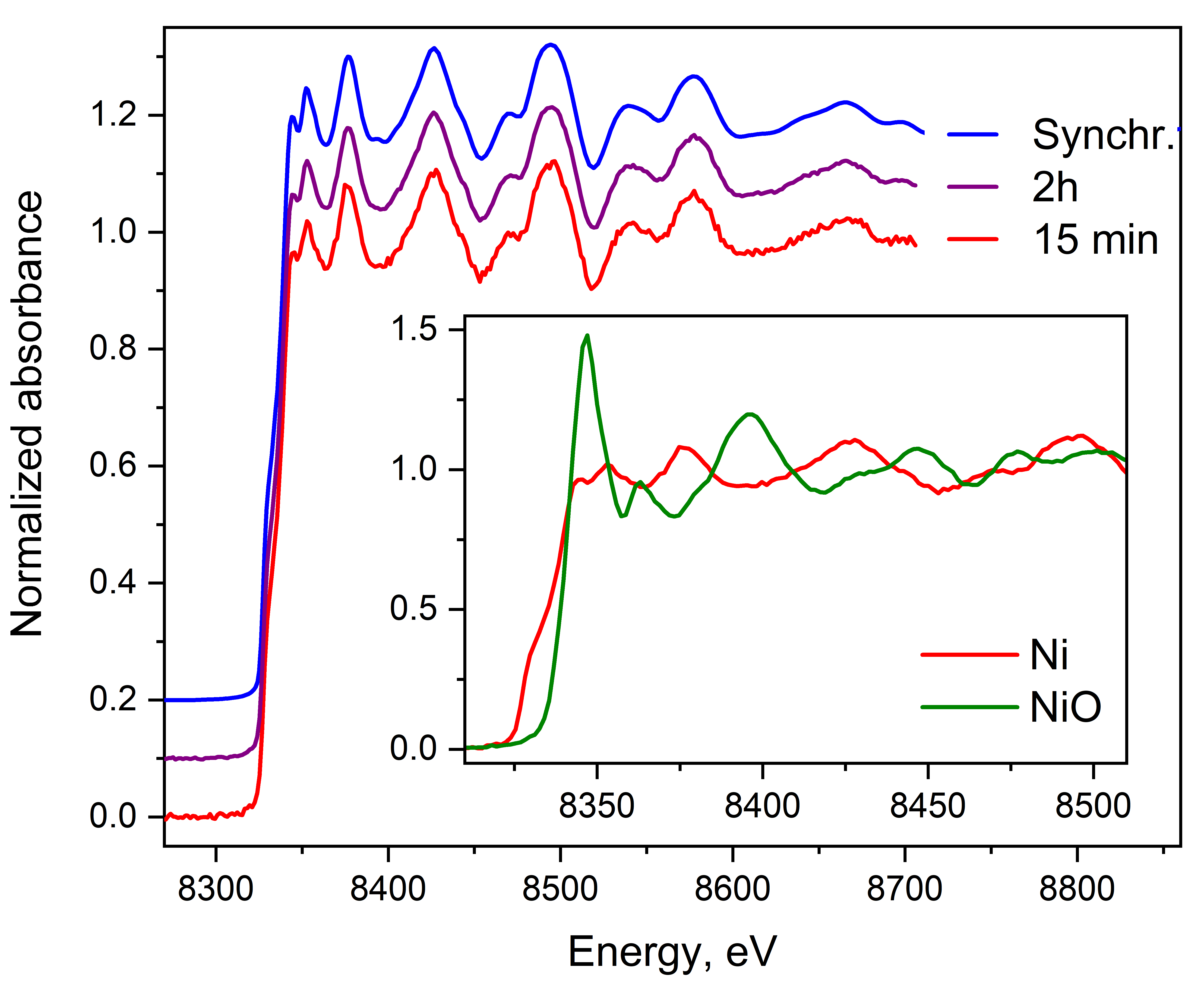}
\end{center}
\caption{XAS spectra of a 6 $\mu$m Ni foil , with 15 min (red) and 2 hours (purple) measurement times. A synchrotron measurement is also shown as a reference (blue) and the patterns are shifted for clarity. The inset compares the 15 min measurement with the spectrum of a NiO sample prepared from powder (green) collected with the same total time.}
\label{fig:figure7}
\end{figure}

The XAS spectrum of a 6 $\mu$m Ni foil was recorded with two different measurement times, namely 15 min and 2h, and the results are shown in Fig. \ref{fig:figure7}. The longer measurement time helps to reduce the noise due to counting statistics but even with the shorter time all essential features are clearly observable. A measurement performed with synchrotron radiation is also included as reference (XAFS spectrum of Nickel. https://doi.org/10.48505/nims.3923). In the inset of Fig. \ref{fig:figure7}, the 15 min measurement is compared with the spectrum of a NiO sample prepared from powder and with the same acquisition time. The differences in the positions and shapes of the absorption edge of Ni are again clearly visible, allowing the analysis of oxidation states and coordinations. 

\subsection{EXAFS of Fe}

\begin{figure}[ht] %
\begin{center}
\includegraphics[width=0.5\textwidth]{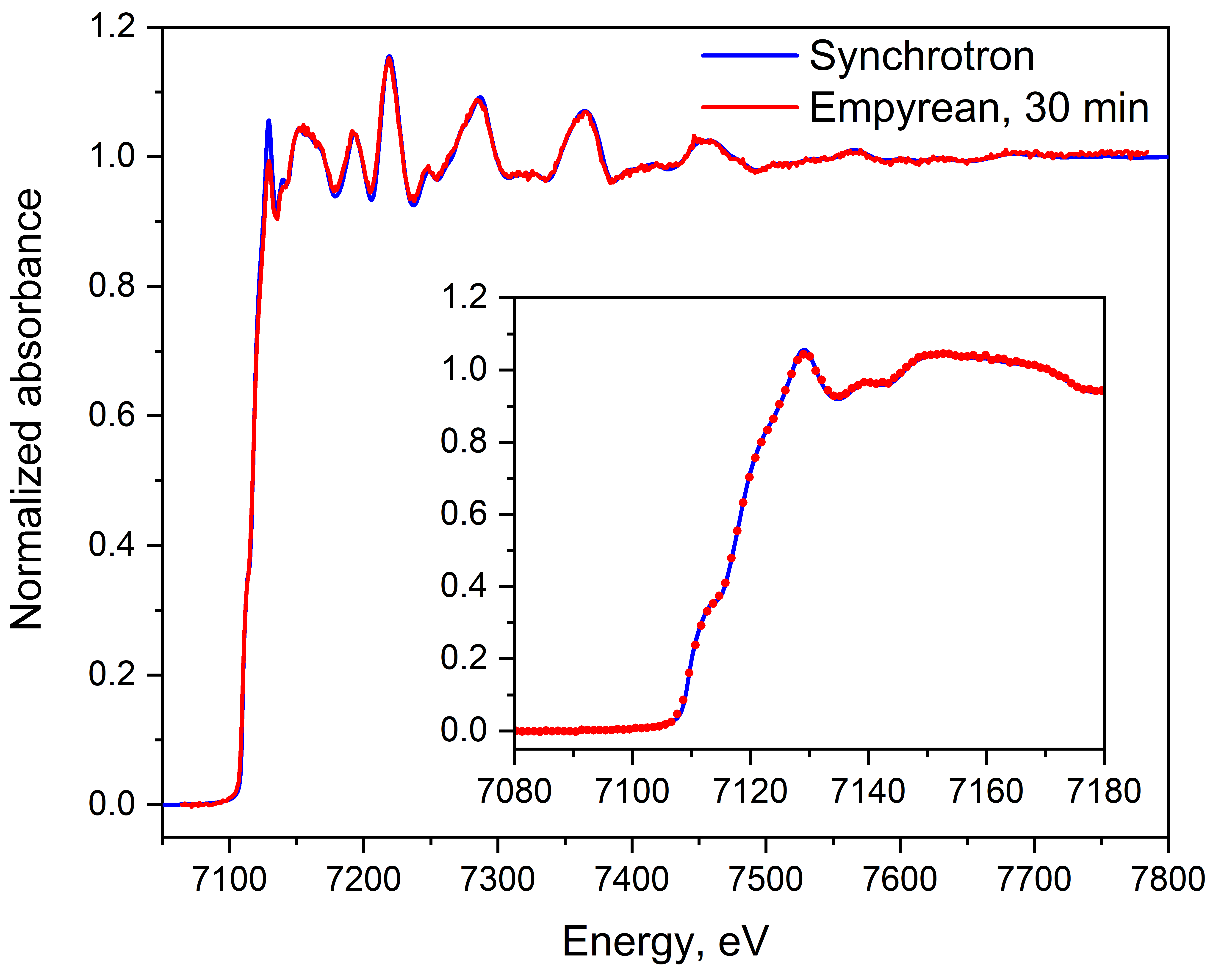}
\end{center}
\caption{Experimental XAS spectrum of a 7.5 $\mu$m Fe foil collected in 30 min and covering a range of 700 eV (red) compared with a synchrotron reference (blue). The inset shows a 30 min scan in the XANES range optimized for better performance.} 
\label{fig:figure8}
\end{figure}

\begin{figure}[ht] %
\begin{center}
\includegraphics[width=0.5\textwidth]{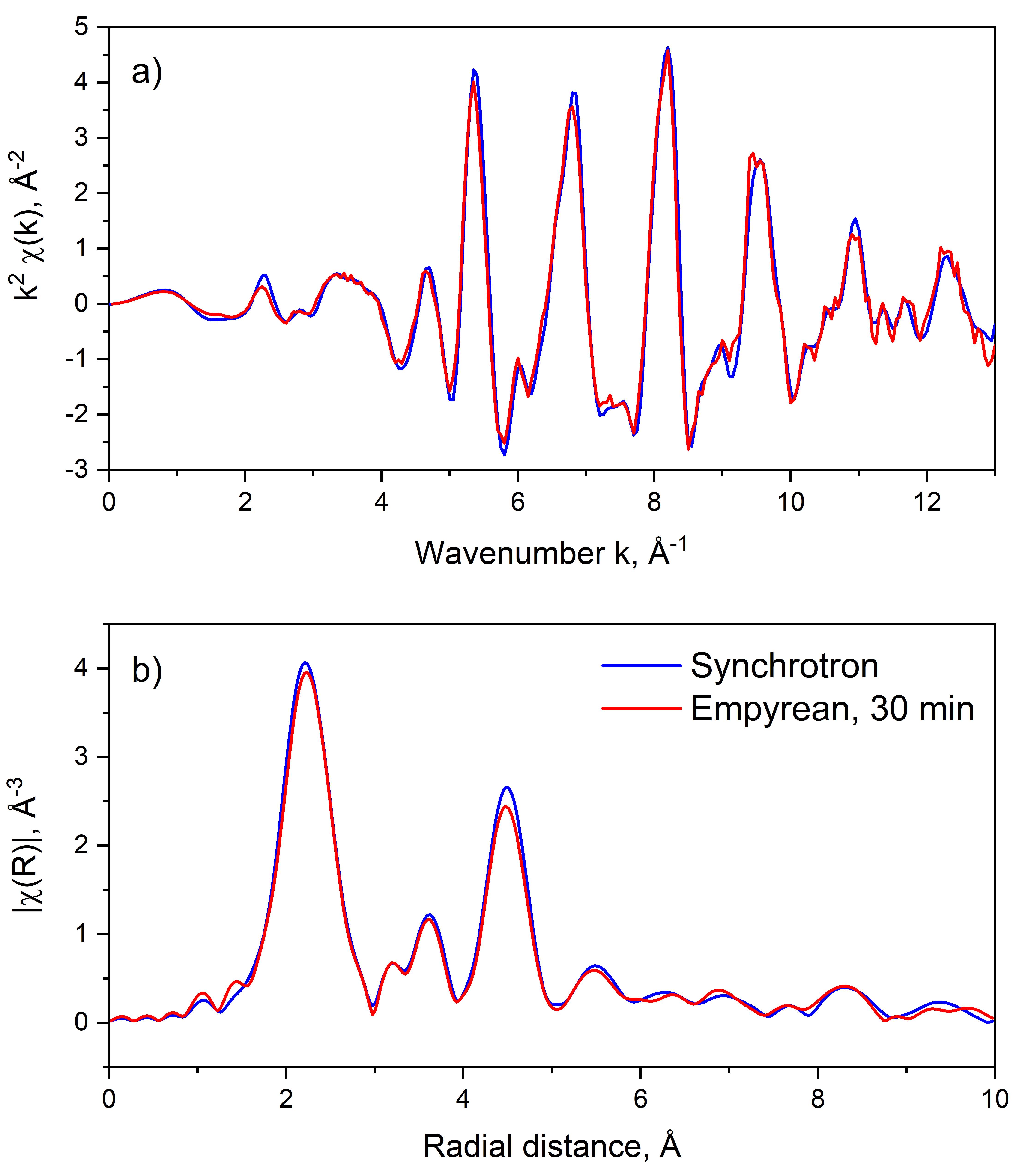} 
\end{center}
\caption{EXAFS signal (a) and the corresponding Fourier transform (b) of Fe calculated from the measurement in Fig. \ref{fig:figure8} (red) compared with the result of the synchrotron measurement adjusted to the same $k$ range (blue).} 
\label{fig:figure9}
\end{figure}

Another XAS spectrum of the 7.5 $\mu$m reference Fe foil is shown in Fig. \ref{fig:figure8}. This measurement covers an energy range of 700 eV that is suitable for EXAFS analysis and was collected in 30 min. In this case, the set-up was optimized for high intensity (with a 1 deg divergence slit), and a small difference with the reference synchrotron measurement (XAFS spectrum of Iron. https://doi.org/10.48505/nims.3903) can be observed close to the absorption edge. Despite this difference, the data can be used to calculate the Fourier transform of the EXAFS signal of the Fe foil and this is very similar to the one calculated from the synchrotron data when the same $k$ range is considered (Fig. \ref{fig:figure9}). After optimization of the set-up for better performance in the near-edge region by using a smaller divergence slit, a second measurement with the same duration and shorter energy range was performed. This measurement is shown in the inset of Fig. \ref{fig:figure8} and is closer to the reference synchrotron data. 

\subsection{EXAFS of Pt L$_3$ edge}

\begin{figure}[ht] %
\begin{center}
\includegraphics[width=0.5\textwidth]{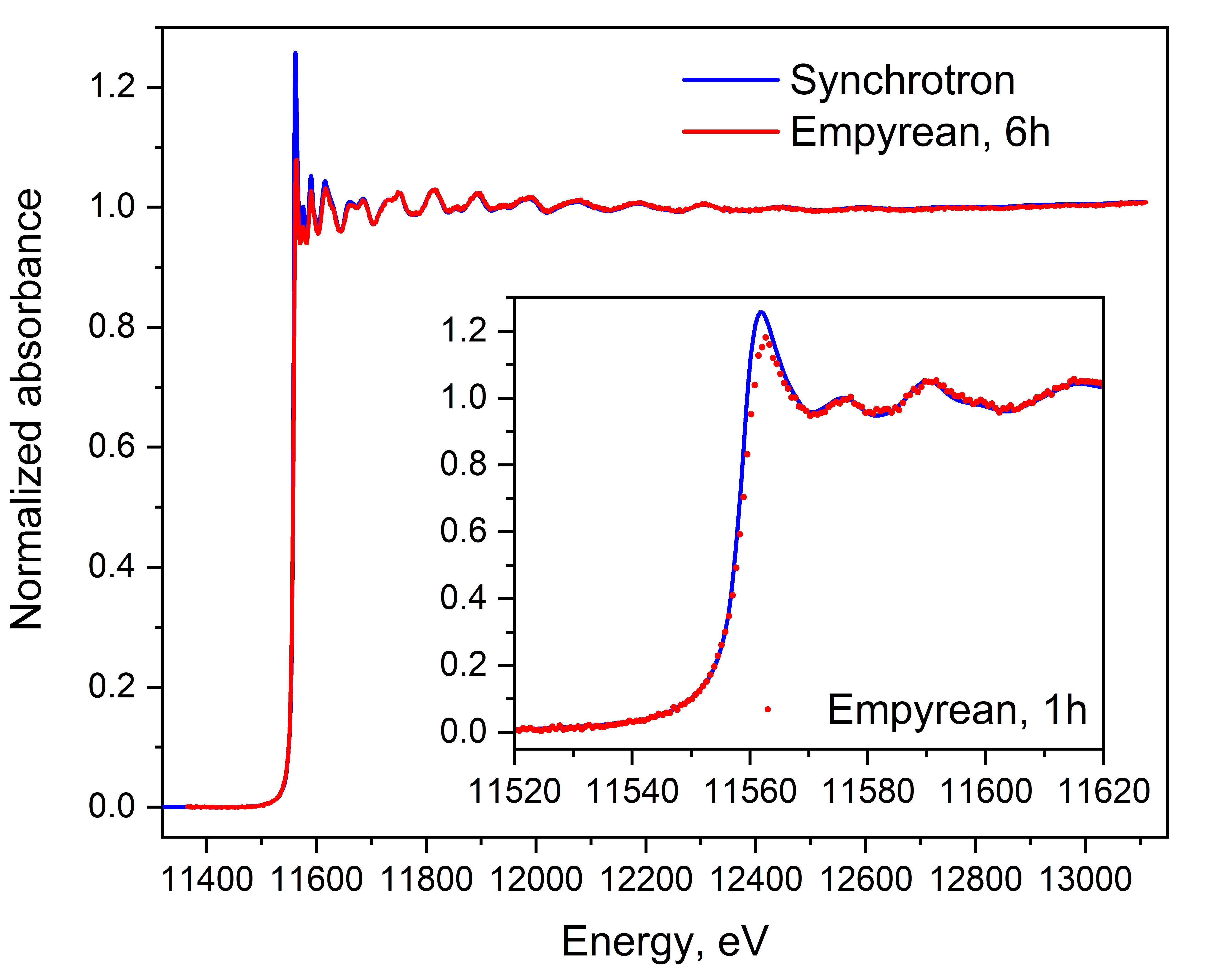} 
\end{center}
\caption{Experimental XAS spectrum of a 7.5 $\mu$m Pt foil, collected in 6 hours and covering a range of 1700 eV (red) compared with a synchrotron reference (blue). The inset shows a 1h scan in the XANES range collected using the Si (044) reflection.  }
\label{fig:figure10}
\end{figure}

\begin{figure}[ht] %
\begin{center}
\includegraphics[width=0.5\textwidth]{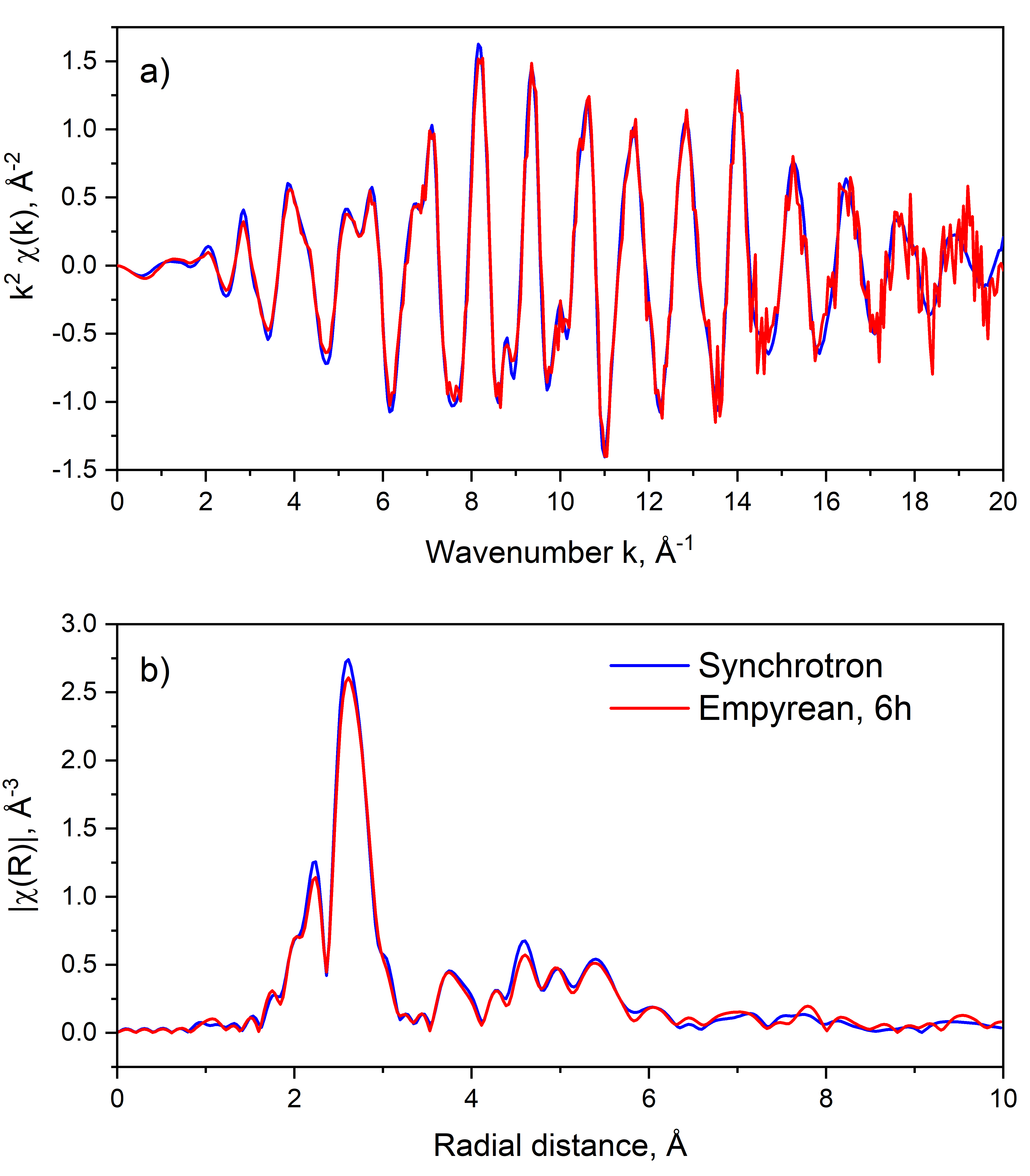}
\end{center}
\caption{EXAFS signal (a) and the corresponding Fourier transform (b) of Pt calculated from the measurement in Fig. \ref{fig:figure11} (red) compared with the result of the synchrotron measurement adjusted to the same $k$ range (blue).}
\label{fig:figure11}
\end{figure}

In order to demonstrate the applicability of the proposed configuration to $L$ absorption edges, the XAS signal from a 7.5 $\mu$m Pt foil was measured in a wide energy range around the $L_3$ edge of Pt. A tube with an Ag anode and power settings 35 kV and 50 mA (see Table \ref{tab:table1}) was used for this measurement, which had a duration of 6 hours. The result is shown in Fig. \ref{fig:figure10}. The laboratory data show good agreement with the synchrotron reference (XAFS spectrum of Platinum. https://doi.org/10.48505/nims.2473), except in the near-edge region where the measured amplitude is lower than the reference. This is likely due to the reduced resolution of the set-up in this energy range. A second measurement with a duration of one hour was conducted using the Si (044) reflection of the same crystal and the outcome is shown in the inset of Fig. \ref{fig:figure10}. With the improved resolution, the features in the XANES region are closer to the reference data. The EXAFS signal of the Pt foil (Fig. \ref{fig:figure11}(a)) agrees well with the synchrotron measurement up to 20 \AA$^{-1}$ in $k$ space. The Fourier transform of the EXAFS signal is also very similar to that calculated from the reference data when the same $k$ range is considered (Fig. \ref{fig:figure11}(b)).

\subsection{XAS of austenitic stainless steel at the Fe, Cr and Ni edges.}

\begin{figure}[ht] %
\begin{center}
\includegraphics[width=0.5\textwidth]{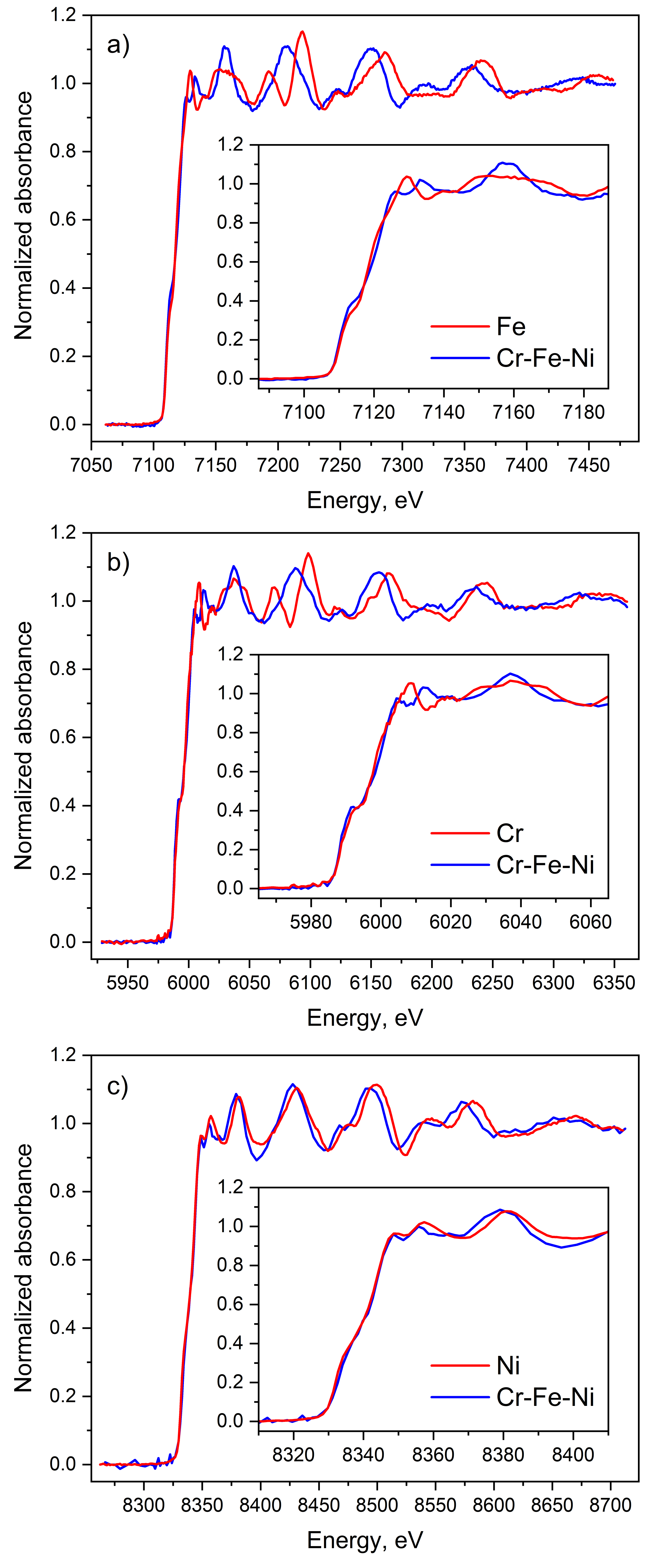} 
\end{center}
\caption{(a) XAS spectrum of a Cr-Fe-Ni alloy with nominal composition 70\% Fe, 16\% Cr, 14\% Ni and a thickness of 5 $\mu$m measured around the absorption edge of Fe, compared with a reference Fe foil with thickness 7.5 $\mu$m. (b) The same alloy measured around the absorption edge of Cr and compared with a reference sample consisting of 1 $\mu$m Cr layer deposited on a 6 $\mu$m Al foil, and (c) around the Ni edge with a reference measurement of a 6 $\mu$m Ni foil. The insets show magnified views of the corresponding XANES ranges.}
\label{fig:figure12}
\end{figure}

The XAS spectrum of the reference Fe foil was also compared with a foil of austenitic steel with nominal composition 70\% Fe, 16\% Cr and 14\% Ni and thickness of 5 $\mu$m. The measurement time for each scan was one hour. While pure Fe crystallizes with the body-centered cubic (bcc) structure, the austenitic steel has a face-centered cubic (fcc) structure at room temperature. The position of the absorption edge is nearly the same in both materials (Fig. \ref{fig:figure12}(a)), however the signal in the EXAFS range is quite different, reflecting the change of the structural type. The same structural change is observed also at the Cr $K$ edge (Fig \ref{fig:figure12}(b)). The collection time for each scan was two hours and the tube settings were adjusted to 30 kV and 55 mA for maximizing the intensity. The Cr reference specimen consists of a 1 $\mu$m Cr layer deposited on 6 $\mu$m Aluminum foil. The measured signal around the Cr edge is significantly weaker and requires longer measurement times compared to those used for the Fe edge. There are several factors that contribute to this. Due to the small amount of Cr atoms in the two samples, the absorbance steps are around 0.3 and are thus significantly smaller than the value that is considered optimal for XAS measurements, which is 1. For the energy corresponding to the Cr absorption edge the transmission through air is only 28\% compared to 47\% for Fe (see Fig. \ref{fig:figure5}). Finally, the absorption by the other elements present in the specimens (Fe, Ni and Al, respectively) also reduce the intensity. The combination of these factors causes that the noise level in the data is more than three times worse for the Cr edge compared to the Fe edge. Nevertheless, the differences between the two Cr-containing samples are clearly observed. Since a high-resolution measurement is not required in the pre-edge and post-edge regions, the step-size of the data shown in Fig. \ref{fig:figure12}(b) was increased by a factor of 3 in these regions by rebinning the measured data. This helps to reduce the visible statistical noise. The near-edge region is not modified. Fig. \ref{fig:figure12}(c) shows the XAS spectrum of the steel sample around the Ni $K$ edge compared with the 6 $\mu$m Ni reference foil. Pure Ni crystallizes with the fcc structure and the two materials have similar features in the EXAFS range. However, for the steel sample the oscillations are shifted to lower energies, indicating a different environment of the Ni atoms. The weight fraction of Ni in the alloy is even lower than that of Cr, and the absorbance step is only 0.17. In addition, photons with this energy are strongly absorbed by the other elements in the material, especially by the Fe atoms. This again results in higher noise levels. The same rebinning procedure was applied to the Ni data as described for the Cr case. 
Despite several experimental factors that affect negatively the collected data, such as low number of absorbing atoms and strong attenuation by the matrix of the material and the air, the measurement set-up was able to provide meaningful XAS results for all three edges that can be used for further analysis in this practical case.

\subsection{\textit{In operando} investigation of an NMC811 battery}

Understanding the electrochemical processes and their relation to the crystalline structures of the cathode and anode materials is important for improving the capacity and lifetime of batteries. \textit{In operando} XAS measurements performed during charging and discharging allow the observation of changes in the oxidation states and local coordinations of different atomic types present in the materials \cite{Jahrman2019b, Genz2024}. A pouch cell battery with an NMC811 cathode (LiNi$_{0.8}$Mn$_{0.1}$Co$_{0.1}$O$_2$) and nominal capacity 50 mAh was investigated in transmission mode with the experimental configuration shown in Fig. \ref{fig:figure1} and using a Si (111) crystal. A constant current - constant voltage (CC-CV) charging strategy was employed in the experiment with 0.2C charging rate using an SP-50e potentiostat from Biologic (Claix, France). The lowest and highest applied voltages were 3.0 and 4.2 V, respectively. The charge-discharge cycle was repeated two times and XAS patterns of the Ni absorption edge were collected during the process. Fig. \ref{fig:figure13} shows a color plot consisting of twenty-two scans, each one with a duration of one hour. In Fig. \ref{fig:figure13}(a) the shift of the absorption edge position between the low and high voltages can be observed, indicating a change of oxidation state of the Ni atoms in the charged and discharged states. The shift is $\sim$2 eV between 3.0 V and 4.2 V. Fig. \ref{fig:figure13}(b) shows the variation of the XANES signal at energies above the absorption edge that can be attributed to the change in the local environment of the Ni atoms. The results  presented here are consistent with the \textit{in operando} data reported in \cite{Kondrakov2017, Tsai2005, Jahrman2019b}. In this case, the X-ray intensity at the Ni edge is attenuated by the battery components, such as the Cu and Al current collectors and the Al pouch.  

\begin{figure}[ht] %
\label{fig:figure13}
\begin{center}
\includegraphics[width=0.5\textwidth]{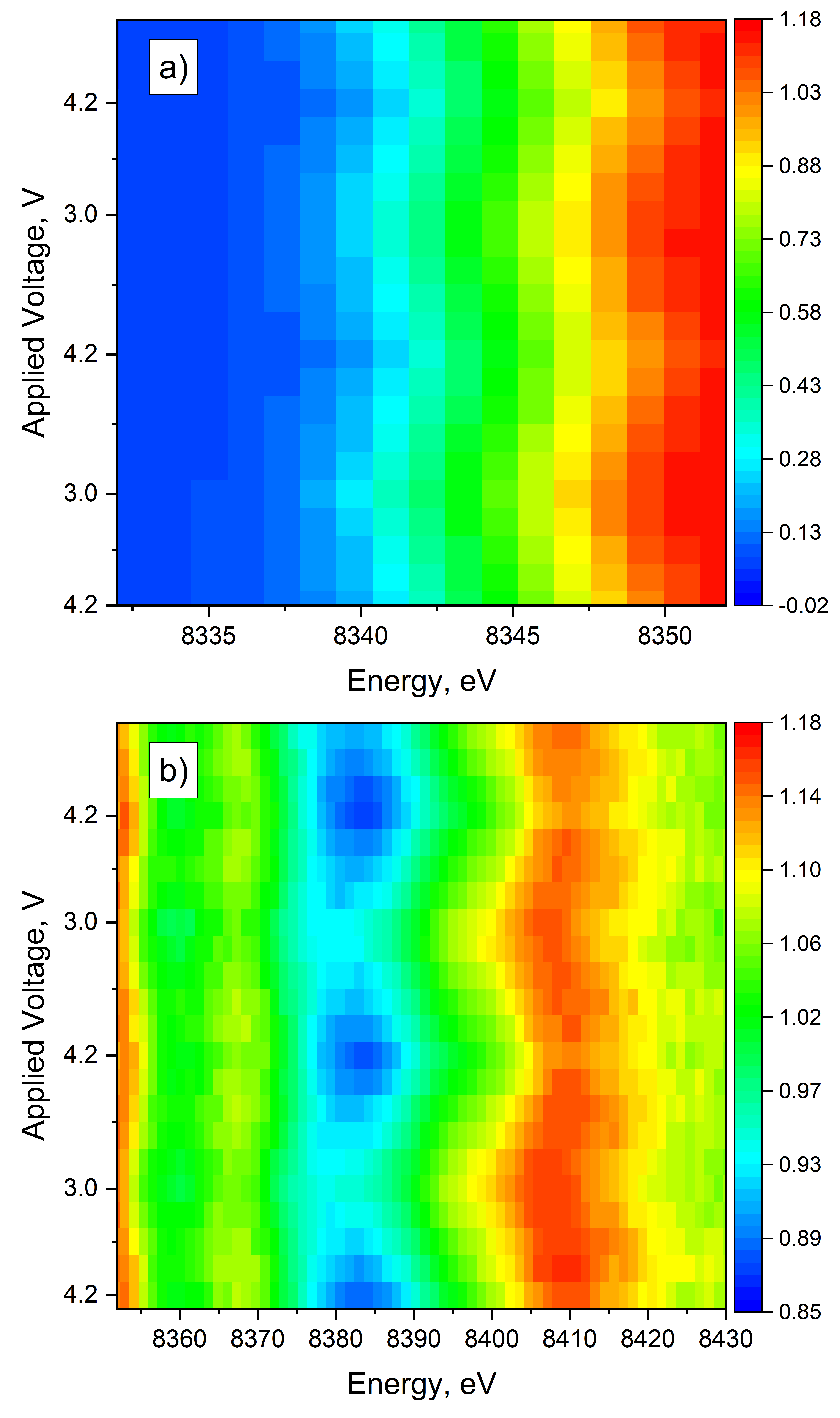}
\end{center}
\caption{2D color plots of the variation of the XAS signal measured during charging and discharging of an NMC811 pouch cell battery: (a) XANES range close to the absorption edge of Ni; (b) XANES range above the edge. The color scheme shows the level of the normalized absorbance.}
\label{fig:figure13}
\end{figure}

\section{Conclusions}

The configuration for XAS measurements presented in the current work is implemented on a standard laboratory powder diffractometer with only minor modifications of the hardware and the control software. It has been tested for a number of different materials and the results show that good data quality can be achieved within a reasonable time, ranging from minutes to hours, depending on the composition of the sample, the sample preparation, the extent of the measurement region and other factors. The main differences between this implementation and other laboratory instruments for XAS are the use of a position-sensitive detector with high energy resolution, very accurate goniometer and the application of continuous scanning of the X-ray source and detector during the measurements. These features are found in modern powder diffractometers, which can be easily reconfigured from diffraction to XAS mode by mounting the crystal analyzer in the center of the goniometer. One advantage of the proposed method is the ability to cover a wide energy range that includes the absorption edges of multiple elements without exchanging the analyzer crystal or other optical components. It also gives the option to switch from high-intensity mode to high-resolution mode by simply repositioning the goniometer to a higher angle corresponding to a different reflection. Enabling XAS measurements with a diffractometer may serve as an entry point for users of diffraction methods who would like to explore XAS and can contribute to the wider use of this technique.

\begin{acknowledgements}
The authors acknowledge prof. Yang-Kook Sun from Hanyang University, Korea for providing the NMC811 cell used in this study. Also, Lei Ding, Sander Weijers, Vladimir Jovanovic and Ferit Cakmak from Malvern Panalytical are acknowledged for their help with the sample preparation and the optimization of the experimental set-up.
\end{acknowledgements}

\ConflictsOfInterest{M.G. declares a patent pending related to the method presented in this work.
}

\DataAvailability{Data sets generated during the current study are available from the corresponding author on reasonable request.
}

\bibliography{XASpaper} 

\end{document}